\documentclass[
 aps,
 pre,
 amsmath,amssymb,
 reprint,
]{revtex4-1}

\usepackage{graphicx}
\usepackage{amsmath}
\usepackage{amssymb}
\usepackage{bm}

\begin{document}

\title{Symmetry-breaking and hysteretic transitions in an actively coupled optical trimer}

\author{Johanne Hizanidis}
\affiliation{
Institute of Electronic Structure and Laser,
Foundation for Research and Technology-Hellas,
70013 Heraklion, Greece
}

\author{Konstantinos G. Makris}
\affiliation{
ITCP--Department of Physics,
University of Crete,
Heraklion, Greece
}
\affiliation{
Institute of Electronic Structure and Laser,
Foundation for Research and Technology-Hellas,
70013 Heraklion, Greece
}

\date{\today}

\begin{abstract}
We investigate the nonlinear dynamics of a non-Hermitian optical trimer consisting of three lossy waveguides with complex-valued coupling. Building on our previous analysis of its global dynamics, we adopt a site-resolved approach to elucidate the bifurcation structure and the organization of oscillatory states across the individual waveguides. We identify a hysteretic transition from reflection-asymmetric periodic states to periodic states exhibiting a combined reflection-translation symmetry. This transition involves a double-loop hysteretic regime characterized by the coexistence of multiple attractors. Analysis of the spatial intensity profiles and their Fourier spectra reveals distinct oscillatory behavior among the waveguides. In particular, in the regime with combined reflection-translation symmetry, the central waveguide exhibits a dominant spatial wavenumber approximately twice that of the edge waveguides. We further identify a third dynamical regime emerging through a pitchfork bifurcation of limit cycles, which breaks the combined reflection-translation symmetry and gives rise to two reflection-related periodic solutions. Despite this transition, the 2:1 relation between the dominant spatial wavenumbers of the central and edge waveguides persists. These results demonstrate how nonlinear interactions, complex coupling, and dissipation shape multistability, bifurcations, and site-dependent oscillations in a minimal non-Hermitian photonic system.
\end{abstract}

\maketitle


\section{Introduction}

Non-Hermitian photonic systems, characterized by the interplay of gain, loss, and complex-valued coupling, have emerged as versatile platforms for exploring unconventional optical phenomena and dynamical behavior beyond the constraints of conservative systems~\cite{EL2018_NATPHYS,PT13,PT14}. Their ability to incorporate dissipative interactions and engineered coupling mechanisms has enabled the observation of exceptional points, unidirectional transport, and parity-time ($\mathcal{PT}$)-symmetry breaking in compact optical configurations~\cite{PT1,PT2,PT3,PT4,PT5,PT6,PT8,PT9,PT10,PT11,PT12,li2013revisiting}. Among the simplest realizations are arrays of coupled optical waveguides, in which light propagates through evanescently coupled channels. The interplay between optical nonlinearity, dissipation, and coupling in these systems gives rise to a variety of dynamical phenomena, including multistability, self-trapping, and chaos~\cite{CHR03,CHR88,JEN82,FIN90,EIL85}.

Within this framework, the actively coupled (AC) dimer, introduced in Ref.~\cite{BAR14}, consists of two lossy waveguides coupled through a complex-valued coefficient whose dissipative component provides an effective gain contribution. This configuration offers an alternative to conventional $\mathcal{PT}$-symmetric systems, for which gain and loss must be balanced between the two components. The extension of this concept to three coupled waveguides, introduced in Ref.~\cite{HIZ24_PRE}, provides a minimal setting in which to investigate collective nonlinear dynamics in non-Hermitian photonic lattices. Despite its simple geometry, the AC trimer exhibits a remarkably rich dynamical structure. In our previous work~\cite{HIZ24_PRE}, we investigated its global response by focusing on the evolution of the total optical power. We identified analytically the stationary solutions and characterized their stability, while numerical simulations and bifurcation analysis revealed the coexistence of oscillatory states, period-doubling sequences, and chaotic dynamics. In particular, the maximum Lyapunov exponent calculated over the coupling--loss parameter plane revealed an intricate arrangement of regular and chaotic regions, with a fractal-like organization.

While the total optical power provides an overall measure of the system's dynamics, it does not fully capture how the nonlinear evolution is distributed among the individual waveguides. In the present work, we extend our previous analysis by adopting a site-resolved perspective, examining the intensity profiles and spatial spectra of the three waveguides separately. This approach reveals a rich dynamical structure involving multistability, chaotic behavior, and hysteretic transitions between distinct periodic regimes. In particular, we identify a transition from reflection-asymmetric oscillations to periodic states exhibiting a combined reflection-translation symmetry. A further transition occurs through a pitchfork bifurcation of limit cycles, which breaks this symmetry and gives rise to two reflection-related periodic solutions. These findings highlight the interplay between nonlinear coupling, dissipation, and reflection-related bifurcations in shaping the dynamics of a minimal non-Hermitian photonic system.

The paper is organized as follows. In Sec.~\ref{sec:1}, we introduce the trimer model and present its bifurcation structure in the coupling-parameter plane, identifying the parameter region selected for the subsequent analysis. In Sec.~\ref{sec:2}, we examine the site-resolved intensity dynamics and characterize the distinct oscillatory regimes. Section~\ref{sec:3} explores the hysteretic transitions and multistability associated with the successive fold bifurcations of limit cycles. In Sec.~\ref{sec:4}, we investigate the spatial profiles and Fourier spectra of the different dynamical regimes, focusing in particular on the combined reflection-translation symmetry in region II and its breaking through a pitchfork bifurcation of limit cycles at the transition to region III. Finally, in the conclusions, we summarize the main findings and discuss
further questions.

\section{Model and Bifurcation Structure in the Coupling-Parameter Plane}
\label{sec:1}

The model employed in this work was introduced in Ref.~\cite{HIZ24_PRE} as a spatial extension of the actively coupled dimer presented in~\cite{BAR14}. It represents a minimal non-Hermitian optical configuration consisting of three evanescently coupled, identical waveguides with complex-valued couplings and uniform loss. Within the framework of coupled-mode theory~\cite{christopoulos2024temporal}, the evolution of the optical field envelopes is governed by the following set of nonlinear equations:
\begin{eqnarray}
   \frac{d{\psi_1}}{dz}&=&-\gamma\psi_1+(ik+\alpha)\psi_2 +i |\psi_1|^2\psi_1, \label{eq:1}\\
   \frac{d{\psi_2}}{dz}&=&-\gamma\psi_2+(ik+\alpha)(\psi_1+\psi_3)+i |\psi_2|^2\psi_2, \label{eq:2}\\
   \frac{d{\psi_3}}{dz}&=&-\gamma\psi_3+(ik+\alpha)\psi_2+i |\psi_3|^2\psi_3. \label{eq:3}
\end{eqnarray}
Here, $\psi_j=R_j e^{i\phi_j}$, with $j=1,2,3$, denotes the complex optical field amplitude in the $j$th waveguide, where $R_j$ and $\phi_j$ are its amplitude and phase, respectively, and $z$ is the propagation distance. The corresponding optical intensity is given by $P_j=|\psi_j|^2=R_j^2$.
The parameter $k$ characterizes the strength of the conservative (reactive) coupling between adjacent waveguides, whereas $\alpha$ controls the dissipative coupling, which introduces an effective gain contribution distributed symmetrically between the coupled waveguides. Together with the uniform loss rate $\gamma$, these parameters determine the balance between conservative energy exchange, gain and dissipation in the trimer, rendering the system non-Hermitian. Consequently, the total optical power, $P(z)=|\psi_1|^2+|\psi_2|^2+|\psi_3|^2$, is no longer conserved and generally varies along the propagation coordinate.

\begin{figure}[]
\includegraphics[width=\columnwidth]{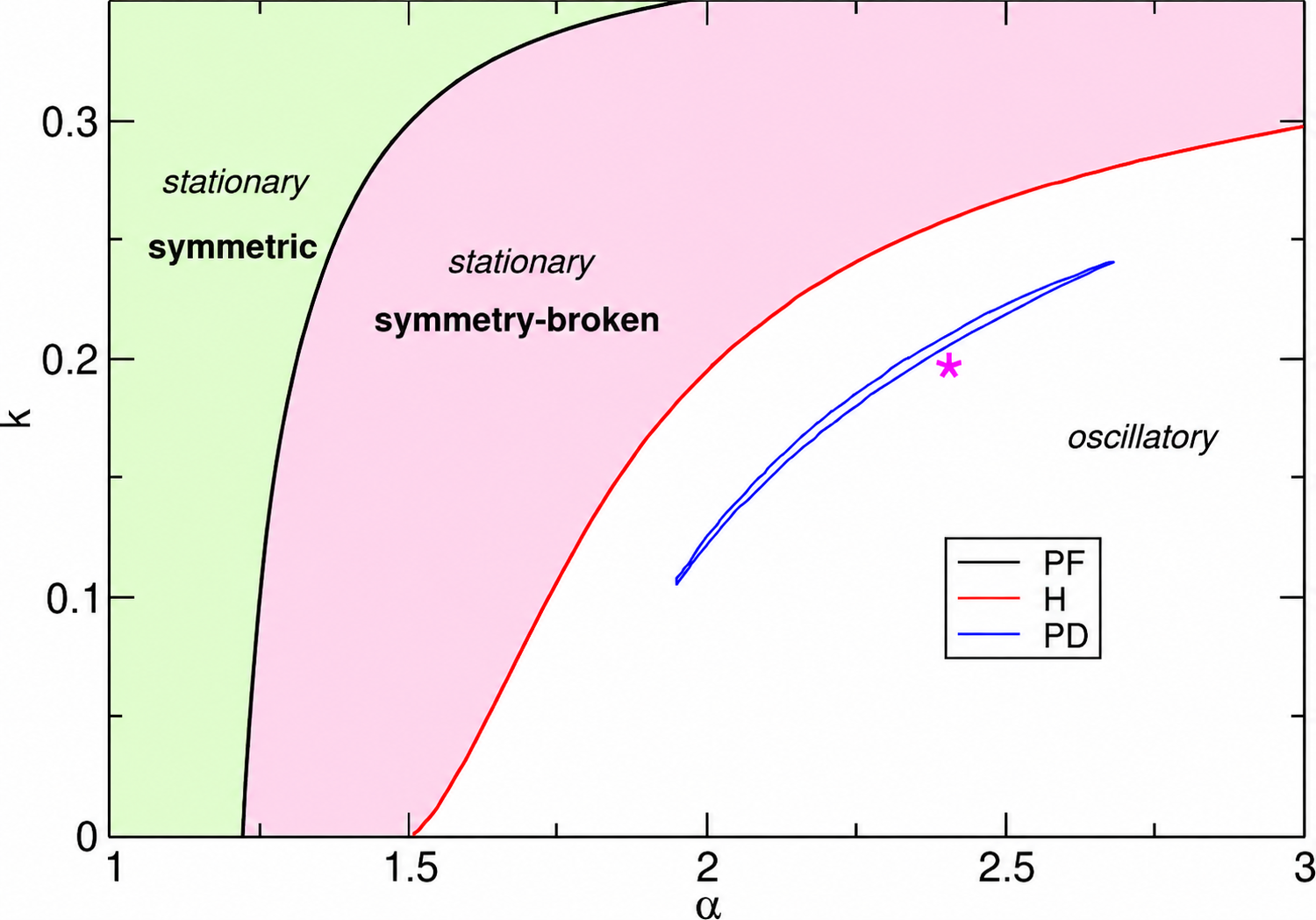}
\caption{Bifurcation diagram in the $(k,\alpha)$ parameter plane for fixed loss $\gamma=1.5$. The pitchfork bifurcation (PF), Hopf bifurcation (H), and period-doubling (PD) curves are shown in black, red, and blue, respectively. The light-green shaded region corresponds to stationary symmetric states, for which $R_1=R_3$, whereas the light-pink region corresponds to stationary symmetry-broken states, for which $R_1\neq R_3$. Beyond the Hopf bifurcation curve, the dynamics becomes oscillatory and exhibits complex behavior, including chaos, as reported in Ref.~\cite{HIZ24_PRE}. The magenta star marks the parameter region where interesting transitions in the oscillatory dynamics occur.
}
\label{fig:1}
\end{figure}

In Ref.~\cite{HIZ24_PRE}, we investigated the dynamical behavior of this trimer from a global perspective, focusing on the collective evolution of the total optical power. That study demonstrated that the system supports analytically tractable stationary states as well as complex oscillatory dynamics, including chaos. In the oscillatory regime, a sequence of bifurcation bubbles and period-doubling bifurcations was identified, eventually leading to chaotic dynamics. The calculation of the maximum Lyapunov exponent across the $(\gamma,\alpha)$ parameter plane revealed a highly intricate arrangement of alternating regular and chaotic regions, exhibiting a fractal-like organization.

Here, we complement that global analysis by examining the bifurcation structure in the $(k,\alpha)$ parameter plane, while keeping the loss rate fixed at $\gamma=1.5$. The resulting bifurcation diagram, shown in Fig.~\ref{fig:1}, provides a complementary view of how the coupling parameters influence the dynamical regimes of the trimer. All numerical simulations presented in this work are performed by integrating the reduced system in polar coordinates, obtained after eliminating the common phase, as described in Appendix~\ref{app:polar}. In particular, although the stationary solutions were already identified in Ref.~\cite{HIZ24_PRE}, their bifurcation structure is further elucidated here by explicitly tracing the pitchfork bifurcation (PF) curve. Likewise, the Hopf bifurcation (H) curve, previously studied in the $(\gamma,\alpha)$ plane, is now continued in the $(k,\alpha)$ plane. The diagram also includes the first period-doubling (PD) bifurcation curve, associated with the onset of the period-doubling route to chaos reported in Ref.~\cite{HIZ24_PRE}. The bifurcation curves are computed using the reduced system, following the continuation procedures and bifurcation conditions detailed in Appendix~\ref{app:bifurcation}.

For clarity, we do not display the multiple period-doubling branches found beyond the first PD bifurcation in Ref.~\cite{HIZ24_PRE}. Instead, Fig.~\ref{fig:1} shows the essential sequence of bifurcations, PF--H--PD, providing the framework for the analysis that follows. Our interest lies in the dynamical regimes beyond the first PD bifurcation, where the global response becomes increasingly complex and the coexistence of distinct attractors gives rise to rich dynamical behavior.

A closer examination of this parameter region reveals features in the spatial organization of the dynamics that cannot be captured by the global description in terms of the total optical power alone. In particular, the individual waveguides can exhibit markedly different intensity profiles, motivating a site-resolved investigation of their dynamics. Although these features occur over an extended portion of the parameter plane beyond the PD curve, we focus on a representative parameter path at fixed $k=0.2$ and $\gamma=1.5$, corresponding to the magenta star in Fig.~\ref{fig:1}. Unless otherwise stated, these parameters are kept fixed throughout the remainder of the paper, while $\alpha$ is varied.

\section{Site-resolved dynamics}
\label{sec:2}
As discussed in the previous section, the bifurcation diagram identifies the parameter region beyond the first period-doubling bifurcation as a particularly interesting regime for further investigation. We now turn to the site-resolved dynamics along the selected parameter path, focusing on the intensity evolution in the individual waveguides and the transitions between the different dynamical regimes.

\begin{figure}[]
\includegraphics[width=\columnwidth]{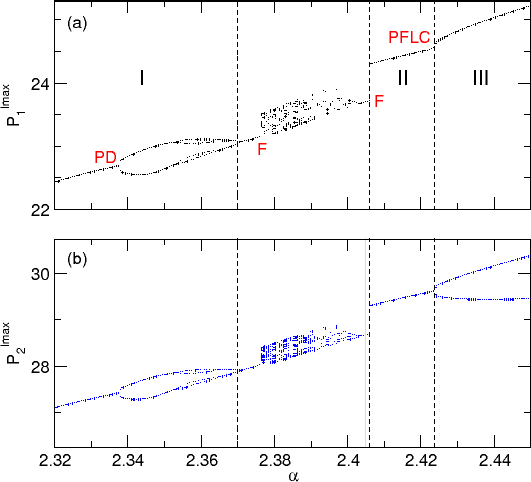}
\caption{Local maxima of the intensities in waveguides 1 and 2 as the parameter $\alpha$ is varied: (a) $P_1$ (black), corresponding to one of the edge waveguides, which are interchangeable, and (b) $P_2$ (blue), corresponding to the central waveguide. The vertical dashed lines delimit the three parameter regions, I--III. The labels F, PD, and PFLC indicate fold bifurcations of limit cycles, period-doubling bifurcations, and the pitchfork bifurcation of limit cycles, respectively.}
\label{fig:2}
\end{figure}

Figure~\ref{fig:2} displays the local maxima of the intensities $P_1=|R_1|^2$ and $P_2=|R_2|^2$ in waveguides 1 and 2, respectively, as functions of the control parameter~$\alpha$. The results are obtained by numerical continuation, using the final state of each simulation as the initial condition for the subsequent one. The labels PD, F, and PFLC indicate the locations of the period-doubling, fold, and pitchfork bifurcations of limit cycles, respectively. Three distinct oscillatory regimes, labeled I, II, and III, can be identified, with regions I and II separated by a narrow parameter interval associated with a hysteretic transition, while region III emerges from region II through a pitchfork bifurcation of limit cycles.

\begin{figure}[]
\includegraphics[width=\columnwidth]{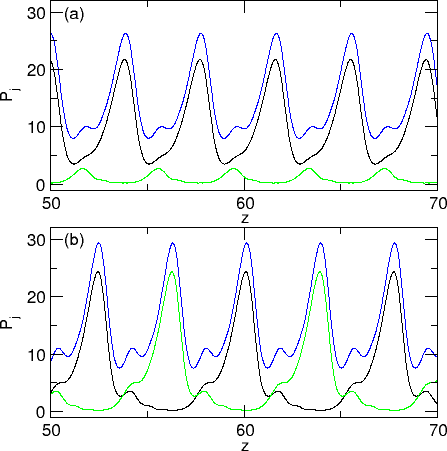}
\caption{Intensity profiles $P_j$ along the propagation coordinate $z$ in the three waveguides for (a) $\alpha=2.33$ (region~I) and (b) $\alpha=2.41$ (region~II). The black, blue, and green curves represent waveguides $j=1$, $2$, and $3$, respectively.}
\label{fig:3}
\end{figure}

In region~I, the oscillatory states are reflection-asymmetric, with unequal intensity profiles in the two edge waveguides. They are the oscillatory counterparts of the asymmetric stationary solutions emerging from the pitchfork bifurcation (PF). A representative solution at $\alpha=2.33$ is shown in Fig.~\ref{fig:3}(a), where the intensity profiles of all three waveguides are displayed. Interestingly, the intensity profiles of the two edge waveguides are shifted relative to one another along the propagation coordinate, while the central-waveguide intensity oscillates approximately in phase with that of the more intense edge waveguide. We refer to these solutions as state~I.

As $\alpha$ increases, the oscillatory branch undergoes a period-doubling (PD) bifurcation, followed by the formation and subsequent disappearance of a period-doubling bubble, eventually leading to chaotic dynamics. This sequence, already reported in Ref.~\cite{HIZ24_PRE}, is not the focus of the present study. Instead, we concentrate on the periodic solutions in region~II, a representative example of which is shown in Fig.~\ref{fig:3}(b) for $\alpha=2.41$.

Unlike state~I, state~II exhibits a combined reflection and translation symmetry along the propagation coordinate. Specifically, the intensity profiles of the two edge waveguides are interchanged by a translation equal to half their spatial period, so that the profile of one edge waveguide coincides with that of the other, and vice versa. The central-waveguide profile, on the other hand, remains invariant under this combined operation, so that its period is half that of the edge waveguides. At the level of the relative phases, the same relation involves the exchange of the two phase differences accompanied by a change of sign. Thus, reflection is recovered only when accompanied by a translation by half the spatial period of the edge-waveguide profiles. The precise conditions are given in Appendix~\ref{app:symmetry}.

At the PFLC, this combined reflection-translation symmetry is broken and two reflection-related periodic solutions emerge. These solutions constitute region~III. Although the intensity profiles of the two edge waveguides remain approximately shifted by half a spatial period, their peak heights are no longer identical. Moreover, the central-waveguide intensity profile is no longer invariant under the same half-period translation. These features distinguish state~III from state~II and are examined in detail in Sec.~\ref{sec:4}.

\section{Double-loop hysteresis}
\label{sec:3}

The transition between states~I and~II involves a rich sequence of dynamical changes, which we examine in detail in this section. As discussed above, state~II exhibits a combined reflection and translation symmetry, in contrast to the reflection-asymmetric state~I. The two states also differ in their intensity levels, with state~II occupying a branch with higher intensity maxima.

\begin{figure*}[]
\includegraphics[width=.9\textwidth]{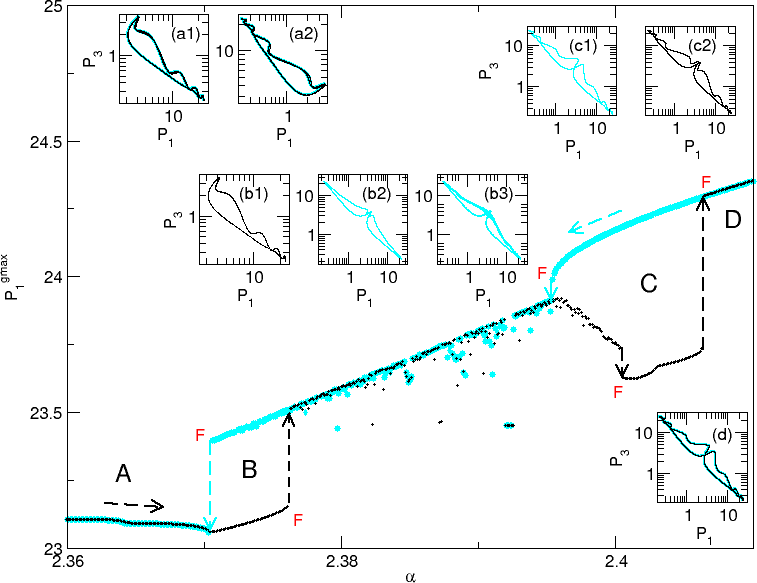}
\caption{Global maxima of $P_1$ versus the control parameter $\alpha$ during the upsweep (black) and downsweep (cyan). Regions A–D illustrate the transition from state-I to state-II solutions through a hysteretic regime. The insets show representative phase portraits: (a1)–(a2) coexisting state-I states in A; (b1)–(b3) coexisting state-I, periodic state-II, and chaotic state-II attractors in B; (c1)–(c2) coexisting periodic state-II solutions in C; and (d) the single periodic state-II solution in D. The labels F indicate fold bifurcations of limit cycles. The same color convention applies to the phase portraits.}
\label{fig:4}
\end{figure*}

To further characterize this transition, Fig.~\ref{fig:4} shows the global maxima of $P_1$, denoted by $P_1^{\mathrm{gmax}}$, as the control parameter $\alpha$ is varied. Black and cyan points correspond to the upsweep and downsweep, respectively. The same color code is used for the corresponding orbits in the phase portraits shown in the insets. The diagram reveals a double-loop hysteretic structure, involving the coexistence of multiple attractors and successive transitions between distinct dynamical states.

Region~A corresponds to state~I, where the upsweep and downsweep branches coincide. Two mirror-image asymmetric solutions coexist, as illustrated in insets (a1) and (a2). In (a2), the solution is the mirror image of that shown in (a1), with the roles of waveguides 1 and 3 interchanged. The dynamics in this region is strictly periodic, with no indication of chaos.

As $\alpha$ increases, the system enters the first hysteretic loop, denoted as region~B. During the upsweep, the asymmetric state followed along the black branch persists until it terminates at a fold bifurcation of limit cycles (F), whereupon the system jumps to the cyan branch, corresponding to state~II. Conversely, during the downsweep, the reverse transition occurs at the fold of the cyan branch. Thus, the first transition from state~I to state~II is triggered by the fold bifurcation of the asymmetric periodic orbit. Within region~B, three distinct attractors coexist: the asymmetric periodic state shown in inset (b1), the periodic state-II solution in (b2), and a chaotic state-II attractor in (b3). The chaotic dynamics in this region, as well as its subsequent development, was discussed in Ref.~\cite{HIZ24_PRE}.

Region~C corresponds to the second hysteretic loop, where bistability persists, but now between two periodic state-II solutions, illustrated in insets (c1) and (c2). These solutions are very similar, differing only slightly in their intensity profiles.

Finally, region~D, which corresponds to state~II, is characterized by the merging of the upsweep and downsweep branches, indicating the disappearance of bistability and the recovery of monostability. Only a single periodic state-II solution remains, as illustrated in inset (d), marking the completion of the transition from state~I to state~II.

Transitions between symmetry-broken and symmetric dynamical states have been reported in other nonlinear optical systems with inherent symmetries, such as the polarization-symmetric bistable model~\cite{KITANO84}, where asymmetric chaotic attractors undergo crisis-induced restructuring and eventually merge into statistically symmetric chaotic states. More broadly, chaotic hysteresis has also been investigated in externally driven nonlinear circuits~\cite{sivaganesh2023emergence}. In contrast, the hysteretic transitions observed in the present non-Hermitian trimer connect reflection-asymmetric states to states exhibiting a combined reflection and translation symmetry. Moreover, this behavior emerges intrinsically through the interplay of Kerr nonlinearity, complex-valued coupling, and dissipation, without external forcing.


\section{Distinct Spatial Dynamics of the Central Waveguide}
\label{sec:4}

Having characterized the transition between regions~I and~II, we now turn to the distinct oscillatory behavior of the central waveguide, focusing first on the wavenumber doubling observed in region~II.
There, the central waveguide oscillates at twice the wavenumber of the edge waveguides. This behavior can be understood better by examining the relative positions of the intensity maxima in Fig.~\ref{fig:3}(b): the maxima of the two edge-waveguide profiles occur alternately, separated by half their spatial period, while the central-waveguide intensity reaches a maximum approximately whenever either edge waveguide does. Thus, the central waveguide exhibits two intensity maxima during one spatial period of the edge-waveguide profiles, giving rise to the observed doubled wavenumber. This doubling is a signature of the combined reflection-translation symmetry of state II: since the central waveguide intensity is invariant under the half-period translation (Eq.~(\ref{eq:combined_symmetry})), its period is half that of the edge waveguides.

\begin{figure}[]
\includegraphics[width=.8\columnwidth]{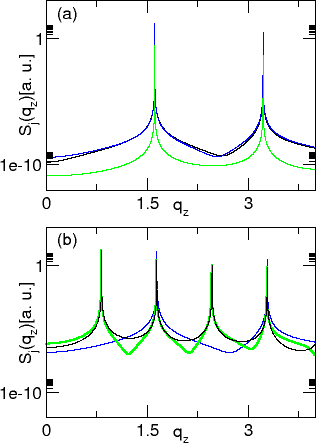}
\caption{Spatial Fourier power spectra for (a) $\alpha=2.33$ (region~I of Fig.~\ref{fig:3}), and (b) $\alpha=2.41$ (region~II of Fig.~\ref{fig:3}). The black, blue, and green curves represent waveguides $j=1$, $2$, and $3$, respectively. The spatial power spectra $S_j(q_z)$ are plotted as functions of the spatial wavenumber $q_z$.}
\label{fig:5}
\end{figure}

To examine this behavior in greater detail, Fig.~\ref{fig:5} presents the spatial Fourier power spectra corresponding to the intensity profiles of Fig.~\ref{fig:3}. Since the system is spatially continuous along the propagation coordinate $z$, but has a discrete transverse structure consisting of three waveguides, we calculate the Fourier power spectrum separately for each waveguide. It is defined as
\begin{equation}
S_j(q_z) =
\left|
\int_{z_0}^{z_0+L}
P_j(z)\,e^{-iq_z z}\,\mathrm{d}z
\right|^2,
\label{eq:power_spectrum}
\end{equation}
where $P_j(z)$ is the intensity in the $j$th waveguide, $L$ is the propagation interval considered, and $q_z$ is the spatial wavenumber associated with variations along the propagation coordinate.

For $\alpha=2.33$ (region~I), all three waveguides oscillate with the same wavenumber, as confirmed by the coinciding dominant peaks in their power spectra. Nevertheless, the intensity profiles exhibit a pronounced amplitude imbalance between the two edge waveguides, characteristic of the reflection-asymmetric state discussed in the previous section. Thus, the three waveguides are wavenumber-synchronized despite their different intensity levels. For $\alpha=2.41$ (region~II), the spectra of the two edge waveguides exhibit coinciding dominant peaks, consistent with their identical intensity profiles up to a spatial translation. The central-waveguide spectrum, however, displays dominant peaks at approximately twice the spatial wavenumber of the edge-waveguide peaks.

\begin{figure*}
    \centering
    \includegraphics[width=\textwidth]{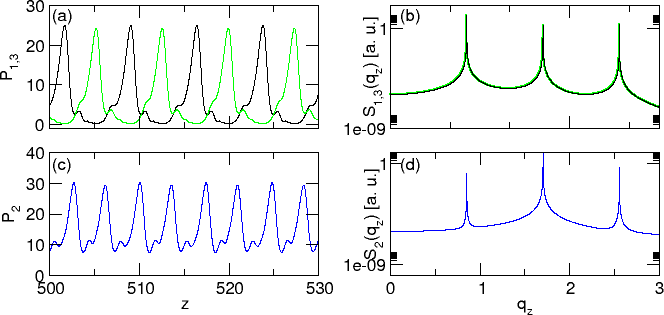}
    \caption{Intensity profiles and corresponding spatial Fourier power spectra at $\alpha=2.44$. (a) Intensity profiles of the edge waveguides, $P_1$ (black) and $P_3$ (green). (b) Corresponding spatial Fourier power spectra, $S_1(q_z)$ and $S_3(q_z)$. (c) Intensity profile of the central waveguide, $P_2$ (blue). (d) Corresponding spatial Fourier power spectrum, $S_2(q_z)$.}
    \label{fig:6}
\end{figure*}

Figure~\ref{fig:6} illustrates the dynamics in regime~III for $\alpha=2.44$, beyond the pitchfork bifurcation of limit cycles (PFLC) marking the transition from state~II to state~III. Panel (a) shows the intensity profiles of the two edge waveguides, $P_1$ (black) and $P_3$ (green), while panel (c) displays the central-waveguide intensity, $P_2$ (blue). Their corresponding spatial Fourier power spectra are shown in panels (b) and (d), respectively.

Following the PFLC, the combined reflection-translation symmetry characterizing state~II is broken. Although the edge-waveguide profiles remain approximately shifted by half their spatial period, their successive peak heights are no longer identical. The central-waveguide intensity also loses its invariance under the same half-period translation, exhibiting a different spatial periodicity. The corresponding Fourier spectra show that the edge-waveguide spectra remain practically identical, while the central-waveguide spectrum exhibits a weaker peak just below $q_z=1$, coinciding with the dominant edge-waveguide wavenumber, and a stronger one just below $q_z=2$. Thus, the dominant wavenumber of the central waveguide remains approximately twice that of the edge waveguides, while the additional lower-wavenumber component reflects the departure from the half-period invariance observed in state~II.

The PFLC gives rise to two periodic solutions related by reflection, which interchange the two edge waveguides while leaving the central waveguide unchanged. The corresponding relations between these solutions, as well as the conditions distinguishing states~II and~III, are discussed in Appendix~C.

\begin{figure}[]
\includegraphics[width=1.0\columnwidth]{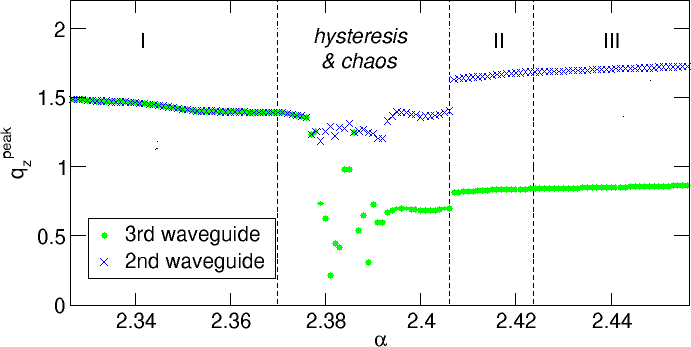}
\caption{Dominant spatial wavenumber $q_z^{\mathrm{peak}}$ extracted from the Fourier power spectra of the waveguide intensities as a function of the control parameter $\alpha$. Green circles represent the edge waveguide, while blue crosses represent the central waveguide. Regions~I, II, and~III correspond to the distinct dynamical regimes identified in the text. The intervening hysteretic and chaotic regime is bounded by the vertical dashed lines.}
\label{fig:7}
\end{figure}

Having established the distinct spectral characteristics of the different regimes, we now examine the evolution of the dominant spatial wavenumbers with the control parameter. Figure~\ref{fig:7} shows the dominant wavenumber, $q_z^{\mathrm{peak}}$, extracted from the Fourier power spectra of the three waveguide intensities as a function of $\alpha$. In region~I, all three waveguides exhibit the same dominant wavenumber, indicating wavenumber synchronization despite the amplitude asymmetry between the edge waveguides. Within the intervening hysteretic and chaotic regime, the dominant wavenumbers vary considerably. In regions~II and~III, the central waveguide exhibits a dominant wavenumber approximately twice that of the edge waveguides, whose wavenumber remains nearly constant. Thus, the 2:1 dominant-wavenumber relation persists across both regions, despite the distinct waveform characteristics of states~II and~III.

\begin{figure*}[]
\includegraphics[width=\textwidth]{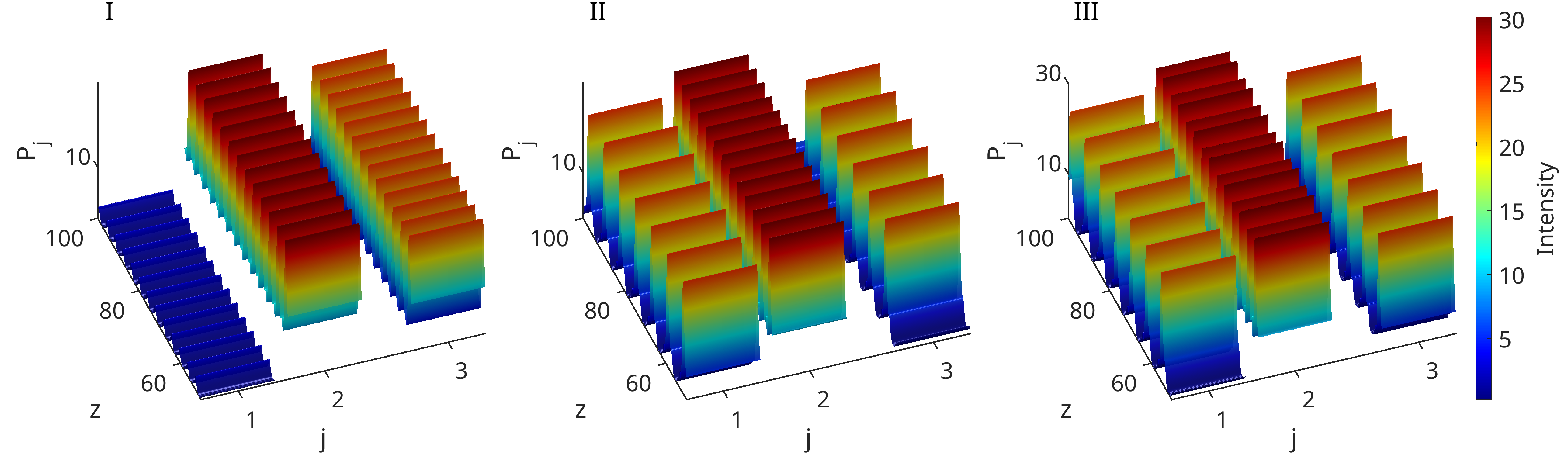}
\caption{Three-dimensional representations of the intensities $P_j$ ($j=1,2,3$) across the three waveguides for representative parameter values: (I) $\alpha=2.33$, (II) $\alpha=2.41$, and (III) $\alpha=2.44$. (I) Reflection-asymmetric periodic state, characterized by unequal edge-waveguide intensities and the same dominant spatial wavenumber in all three waveguides. (II) Periodic state exhibiting combined reflection-translation symmetry, with the edge-waveguide intensity profiles interchanged by a half-period translation and the central-waveguide intensity repeating twice as frequently. (III) Periodic state in which the combined symmetry is broken: the edge-waveguide profiles remain approximately related by a half-period translation, but their peak heights differ slightly, while the central-waveguide intensity exhibits alternating peak heights. The dominant spatial wavenumber of the central waveguide remains approximately twice that of the edge waveguides.}
\label{fig:8}
\end{figure*}

To provide a visual overview of the different dynamical behaviours investigated in this work, Fig.~\ref{fig:8} presents three-dimensional representations of the intensity profiles $P_j(z)$ across the three waveguides for representative parameter values in regions~I, II, and~III, respectively. The side-by-side representations highlight the distinct spatial organization of the oscillations in each regime, bringing together the reflection-asymmetric dynamics of region~I, the combined reflection-translation symmetry of region~II, and its breaking in region~III.

\section{Conclusions}

In this work, we have investigated the nonlinear dynamics of a non-Hermitian optical trimer with complex-valued coupling, extending our previous analysis of its global bifurcation structure to a site-resolved description. By examining the evolution of the individual waveguide intensities, we have identified how the collective dynamics are distributed across the three sites and how their responses are interconnected. More generally, the role of symmetry properties in organizing the bifurcation structure and stability of periodic solutions has been explored in other nonlinear dynamical systems as well~\cite{Marques2004,Crawford1991}.

Our results reveal a rich sequence of dynamical regimes, encompassing reflection-asymmetric periodic oscillations, chaotic dynamics, and periodic states with a combined reflection-translation symmetry. The transition between the first two periodic regimes involves a double-loop hysteretic interval, where multiple attractors coexist and successive transitions occur through fold bifurcations of limit cycles. A further transition from region~II to region~III takes place through a pitchfork bifurcation of limit cycles (PFLC). This bifurcation breaks the combined reflection-translation symmetry of region~II and gives rise to two periodic solutions related by reflection. Although the dominant wavenumber of the central waveguide remains approximately twice that of the edge waveguides, the spatial intensity profiles change significantly across this transition. These findings demonstrate how multistability, hysteresis, chaos, and bifurcations organize the site-dependent dynamics of a minimal non-Hermitian system.

Beyond the specific trimer considered here, these results may provide insight into the dynamics of larger non-Hermitian photonic arrays and other nonlinear oscillator networks. Future work could investigate how parameter inhomogeneity, higher-order couplings, and network topology affect the emergence and robustness of such bifurcations and reflection-related states. More broadly, the interplay between bifurcation structure, reflection relations, and dissipation offers a useful perspective for understanding complex dynamics in coupled nonlinear systems.
\begin{acknowledgments}

This project was funded by the European Research Council
(ERC--Consolidator) under Grant Agreement
No.~101045135 (\emph{Beyond Anderson}).

\end{acknowledgments}

\section*{Data Availability Statement}

The data that support the findings of this study are available
from the corresponding author upon reasonable request.

\appendix
\label{appendix}

\section{Polar-coordinate formulation}
\label{app:polar}

By introducing polar coordinates,
\begin{equation}
\psi_j=R_j e^{i\phi_j}, \qquad j=1,2,3,
\end{equation}
where $R_j$ and $\phi_j$ denote the amplitude and phase of the electric field in the $j$th waveguide, respectively, and defining the two phase differences
\begin{equation}
\theta_1=\phi_2-\phi_1,\qquad
\theta_2=\phi_3-\phi_2,
\end{equation}
the dynamics can be described in terms of the three amplitudes
$R_1,R_2,R_3$ and the two phase differences $\theta_1,\theta_2$.
Substituting this polar representation into the coupled-mode equations,
Eqs.~(\ref{eq:1})--(\ref{eq:3}), and separating real and imaginary parts
yields the following five-dimensional system:
\begin{align}
\frac{dR_1}{dz}={}&-\gamma R_1
+\alpha R_2\cos\theta_1
-k R_2\sin\theta_1,
\label{eq:polar1}\\
\frac{dR_2}{dz}={}&-\gamma R_2
+\alpha R_1\cos\theta_1
+k R_1\sin\theta_1
\nonumber\\
&+\alpha R_3\cos\theta_2
-k R_3\sin\theta_2,
\label{eq:polar2}\\
\frac{dR_3}{dz}={}&-\gamma R_3
+\alpha R_2\cos\theta_2
+k R_2\sin\theta_2,
\label{eq:polar3}\\
\frac{d\theta_1}{dz}={}&R_2^2-R_1^2
-\frac{R_2}{R_1}
\left(\alpha\sin\theta_1+k\cos\theta_1\right)
\nonumber\\
&+\frac{R_1}{R_2}
\left(-\alpha\sin\theta_1+k\cos\theta_1\right)
\nonumber\\
&+\frac{R_3}{R_2}
\left(\alpha\sin\theta_2+k\cos\theta_2\right),
\label{eq:polar4}\\
\frac{d\theta_2}{dz}={}&R_3^2-R_2^2
-\frac{R_2}{R_3}
\left(\alpha\sin\theta_2-k\cos\theta_2\right)
\nonumber\\
&-\frac{R_1}{R_2}
\left(-\alpha\sin\theta_1+k\cos\theta_1\right)
\nonumber\\
&-\frac{R_3}{R_2}
\left(\alpha\sin\theta_2+k\cos\theta_2\right).
\label{eq:polar5}
\end{align}

Stationary solutions of this reduced system correspond to states with
constant amplitudes and constant phase differences. The individual
phases, however, may evolve along the propagation coordinate with a
common phase velocity $\omega$. In the original complex-amplitude
formulation, such solutions take the form
\begin{equation}
\psi_j(z)=R_j e^{i(\omega z+\phi_j)},
\end{equation}
where $R_j$ and $\phi_j$ are constants. Thus, stationary solutions of
the reduced system correspond to uniformly rotating states in the
original formulation.

Similarly, periodic orbits of the reduced system correspond to
modulated rotating states, which in general evolve on invariant tori
in the original complex-amplitude formulation. The bifurcation
analysis below is therefore performed in the reduced system, where
these states are represented by fixed points and periodic orbits.

\section{Bifurcation conditions and numerical continuation}
\label{app:bifurcation}

The bifurcation analysis is performed using the five-dimensional
system in Eqs.~(\ref{eq:polar1})--(\ref{eq:polar5}). The reflection
symmetry breaking of stationary solutions is associated with a
pitchfork bifurcation (PF), at which a real eigenvalue of the
Jacobian matrix crosses zero. The corresponding condition is
\begin{equation}
\lambda_{\mathrm{PF}}=0,
\label{eq:PF_condition}
\end{equation}
where $\lambda_{\mathrm{PF}}$ denotes the eigenvalue associated with
the symmetry-breaking perturbation. The bifurcation gives rise to
two asymmetric branches corresponding to mirror-image solutions.

A Hopf bifurcation (H) occurs when a complex-conjugate pair of
eigenvalues crosses the imaginary axis:
\begin{equation}
\lambda_{1,2}=\pm i\omega_{\mathrm{H}},
\qquad \omega_{\mathrm{H}}\neq 0,
\label{eq:Hopf_condition}
\end{equation}
where $\omega_{\mathrm{H}}$ is the Hopf frequency. This bifurcation
generates periodic orbits of the reduced system.

The periodic orbits may subsequently undergo period-doubling (PD),
fold bifurcations of limit cycles, or pitchfork bifurcations of
limit cycles. A period-doubling bifurcation occurs when a Floquet
multiplier crosses the unit circle through $-1$:
\begin{equation}
\mu_{\mathrm{PD}}=-1.
\label{eq:PD_condition}
\end{equation}
The resulting periodic orbit has twice the spatial period of the
original orbit.

A fold bifurcation of limit cycles occurs when two periodic orbits
coalesce and disappear. It is characterized by an additional
nontrivial Floquet multiplier reaching $+1$:
\begin{equation}
\mu_{\mathrm{F}}=+1.
\label{eq:fold_condition}
\end{equation}
Here, the trivial unit multiplier associated with the
autonomous nature of the periodic orbit is excluded.

A pitchfork bifurcation of limit cycles (PFLC) also occurs when an
additional nontrivial Floquet multiplier reaches $+1$:
\begin{equation}
\mu_{\mathrm{PFLC}}=+1.
\label{eq:PFLC_condition}
\end{equation}
In contrast to a fold bifurcation, this instability is associated
with a symmetry-breaking perturbation of the parent periodic orbit.
In the present system, the relevant symmetry is the combined
reflection and half-period spatial translation. At the bifurcation,
the critical Floquet eigenfunction breaks this symmetry, giving
rise to two symmetry-related periodic orbits that no longer
possess the symmetry of the parent orbit. The two emerging
solutions are mapped onto each other by the reflection operation.

The bifurcation curves shown in Fig.~\ref{fig:1} are obtained by
numerical continuation of the corresponding stationary solutions
and periodic orbits, imposing the appropriate conditions
Eqs.~(\ref{eq:PF_condition})--(\ref{eq:PD_condition}) for each
bifurcation~\cite{Kuznetsov2023}. The numerical continuation and stability calculations
are performed using the methods described in
Refs.~\cite{Sieber2014,Engelborghs2002}.
\section{Reflection and reflection--translation symmetry}
\label{app:symmetry}

The reflection symmetry of the trimer, which exchanges the two edge
waveguides while leaving the central waveguide unchanged, is expressed
in the reduced variables as
\begin{equation}
(R_1,R_2,R_3,\theta_1,\theta_2)
\longrightarrow
(R_3,R_2,R_1,-\theta_2,-\theta_1).
\label{eq:reflection}
\end{equation}
For stationary solutions, reflection symmetry therefore requires
\begin{equation}
R_1=R_3,\qquad \theta_1=-\theta_2.
\label{eq:symmetry_condition}
\end{equation}

We now consider the periodic solutions discussed in the main text.
In state~I, the periodic solutions are reflection-asymmetric, satisfying
\begin{equation}
R_1(z)\neq R_3(z),\qquad
\theta_1(z)\neq-\theta_2(z).
\label{eq:asymmetric_periodic}
\end{equation}

In state~II, the solutions exhibit a combined reflection-translation
symmetry. Specifically, the reflection operation is accompanied by a
translation $\Delta z$ along the propagation coordinate, such that

\begin{equation}
\begin{aligned}
R_1(z+\Delta z)&=R_3(z),\\
R_3(z+\Delta z)&=R_1(z),\\
R_2(z+\Delta z)&=R_2(z),\\
\theta_1(z+\Delta z)&=-\theta_2(z),\\
\theta_2(z+\Delta z)&=-\theta_1(z).
\end{aligned}
\label{eq:combined_symmetry}
\end{equation}
The phase relations are understood modulo $2\pi$. For the solutions
considered here, $\Delta z=T_{\mathrm{edge}}/2$, where
$T_{\mathrm{edge}}$ denotes the spatial period of the edge-waveguide intensity profiles.

In state~III, the combined reflection-translation symmetry characterizing state~II is broken through a pitchfork bifurcation of limit cycles. Although the edge-waveguide intensity profiles remain approximately shifted by half their spatial period, their peak heights are no longer identical. Furthermore, the central-waveguide intensity profile is no longer invariant under the same half-period translation.

To establish the relation between the two solutions emerging in state~III, we consider two initial conditions, denoted by A and B. Here, A denotes the original initial condition, while B denotes the initial condition obtained by applying the reflection transformation to A (Eq.~\ref{eq:reflection}). The resulting solutions satisfy
\begin{equation}
\begin{aligned}
R_1^B(z)&=R_3^A(z),\\
R_2^B(z)&=R_2^A(z),\\
R_3^B(z)&=R_1^A(z),\\
\theta_1^B(z)&=-\theta_2^A(z),\\
\theta_2^B(z)&=-\theta_1^A(z).
\end{aligned}
\label{eq:reflection_related_solutions}
\end{equation}
The relations above hold for any orbit and therefore do not, by themselves, distinguish regions II and III. In region II, solutions A and B coincide up to the spatial translation \(\Delta z\), so that they represent the same periodic orbit. In region III, by contrast, A and B are distinct periodic orbits related by reflection. Thus, the emergence of two distinct reflection-related orbits characterizes the transition to region III.
\end{document}